# Photon emission gain in Er doped Si light emitting diodes by impact excitation

Huayou Liu[1], Jiayuan Zhao[1], Jing Zhang[1], Huan Liu[1], Jiajing He[2], Ulrich Kentsch[3], Shengqiang Zhou[3], Manfred Helm[3], Yaping Dan[1,*]

[1]State Key Laboratory of Advanced Optical Communication Systems and Networks, University of Michigan–Shanghai Jiao Tong University Joint Institute, Shanghai Jiao Tong University, Shanghai 200240, China
[2]Aerospace Laser Technology and System Department, Shanghai Institute of Optics and Fine Mechanics, Chinese Academy of Sciences, Shanghai 201800, China
[3]Institute of Ion Beam Physics and Materials Research, Helmholtz-Zentrum Dresden-Rossendorf, Bautzner Landstr. 400, Dresden, 01328 Germany

Abstract
This work demonstrates photon emission gain, i.e., emission of multiple photons per injected electron, through impact excitation in Er-doped silicon light-emitting diodes (LEDs). Conventional methods for exciting Er ions in silicon suffer from low efficiency due to mismatched energy transfer between exciton recombination and Er excitation. Here, we propose a reverse-biased Si PN junction diode where ballistically accelerated electrons induce inelastic collisions with Er ions, enabling tunable excitation via electric field modulation. Theoretical modeling reveals that photon emission gain arises from multiple impact excitations by a single electron traversing the electroluminescence region, with the gain value approximating the ratio of emission region width to electron mean free path, i.e., $G = L_{ex}/\ell$. Experimental results show an internal quantum efficiency (IQE) of 1.84% at 78 K, representing a 20-fold enhancement over room-temperature performance. This work provides a critical foundation for on-chip integration of silicon-based communication-band lasers and quantum light sources.

An electronically pumped silicon light source emitting single photons is crucial to the implementation of a fully integrated silicon photonic quantum system[1-3]. Er ions implanted in silicon provide an unprecedented opportunity to develop such a system due to the fact that Er ions have a long spin coherent time[4, 5] and emit photons at communication wavelength[6, 7]. Er ions in silicon are often excited by the quantized energy transferred from the recombination of excitons via Er-related defects in the Si band gap[8, 9]. Unfortunately, the quantized transfer energy does not match the Er excitation energy from ground state to the first excited state. As a result, Er ions in Si excited by the recombination of excitons have an extremely low efficiency of photon emission[10-14].

Impact excitation provides a new avenue for exciting Er ions in Si. In a reversely biased Si PN junction diode, electrons are ballistically accelerated within the mean free path to excite Er ions by inelastic collisions. This impact excitation can provide tunable energies by the electric field in the depletion region. As a result, Er doped Si light emitting diodes (LEDs) under reverse bias were previously reported to have much higher luminescence intensities than under forward bias[15-18]. Recently, we established an impact excitation theory to analytically describe the excitation process[15]. The experimental photon emission rate and internal quantum efficiency can be well fitted with the analytical theory, from which we extracted parameters such as the electron mean-free path and the concentration of optically active Er ions that are consistent with literature reports[8, 19].

In this Letter, we report the photon emission gain by impact excitation in Er doped Si LEDs. This can be regarded as analogue (but inverse) to the photoconductive gain in photodetectors[20]. In photodetectors it refers to the generation of multiple electrons form one photon, whereas here we have the generation of multiple photons from one electron, reflected in a superlinear photon vs. electron flux characteristics. By fitting the experimental photon emission with our impact excitation theory, we find that the emission gain is equal to the ratio of the electroluminescence region width to the electron mean-free path. This correlation remains valid for different Er doped Si LEDs operating from 300 K to 77 K. Clearly, the emission gain is induced by the multiple collisions of the hot electrons with Er ions as the electrons transport through the long electroluminescence region.

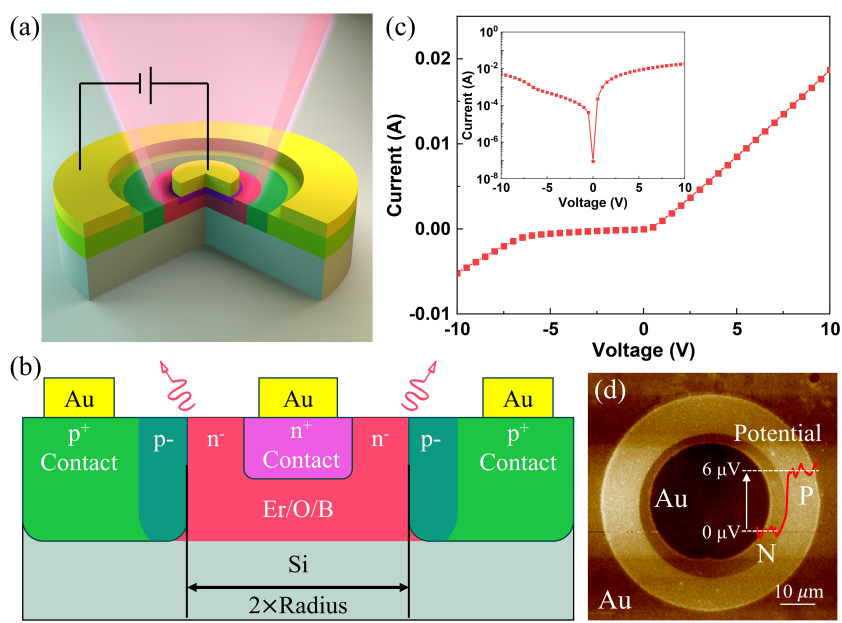

Figure 1. (a) Schematic of Er doped Si LED. (b) Cross section view of the LED. (c) I-V curves of the LED. (d) Surface potential distribution of the device by KPFM.

The structure of the Er-doped silicon LED is schematically shown in **Error! Reference source not found.**(a). To increase the emission gain by impact excitation, we need a wide depletion region in the PN junction diode. According to the device principle of PN junction diodes, reducing the doping concentration of the p and n region will increase the depletion region width. It is well known that Er/O co-doping in Si creates degenerate n-type doping[8, 9, 15]. For this reason, we introduce B dopants into the n-type Er/O region to reduce the electron concentration[15]. Luckily, the introduction of B dopants also helps improve the optical activation of Er ions[8]. To make it easier to tune the electron concentration, we create a uniform doping layer of Er, O and B by using multiple implantation energies and doses. After a systematic tuning of dopant concentration, we have managed to create an n-type Er/O/B layer with the electron concentration of $3.4 \times 10^{17}$ cm$^{-3}$ and a p-type Er/O/B layer with the hole concentration between $10^{15}$ and $10^{16}$ cm$^{-3}$. Simultaneously, the annealing condition is also optimized by adjusting the annealing temperature and time to ensure strong emission from Er ions. The doping profile of Er/O/B, Hall effect measurements, four probe characterizations and Er emission spectra are presented in Supplementary Information (SI) section I. The final doping structure in the device cross-section is shown in **Error! Reference source not found.**(b). **Error! Reference source not found.**(c) shows the current vs voltage (IV) characteristics of a typical Er doped Si LED. **Error! Reference source not found.**(d) is the spatial distribution of the surface electric potential obtained by Kelvin Probe Force Microscopy (KPFM, Dimension XR, Bruker). The potential transition (red line) occurs right at the boundary between intrinsic region and n⁻ region as we designed.

A high-speed short-wave infrared camera (IH320, InGaAs focal plane, Tianying Optoelectronics, China) is used to capture the near infrared light emission from the LED. Figure 2(a) is the optical microscopic image of the LED device under visible light, where the two yellow rings represent the borders of the inner and outer gold electrode. Previously, we found that the Er doped Si LEDs exhibit a much higher emission efficiency under reverse bias than forward bias[15]. An impact excitation theory was established for such high emission efficiency under reverse bias. Figure 2(b) and (c) are the near infrared electroluminescent images for the LED at reverse bias of -9 V and -39 V. The electroluminescence spectrum of the device is shown as the inset in Figure 2(c). By comparing with the KPFM image in **Error! Reference source not found.**(d), we identify the electroluminescence coming from the boundary of the p⁻ and n⁻ region where the depletion region is located. The spatial profiles of the electroluminescence along the red and blue dashed line in Figure 2(b) and (c) are plotted in Figure 2(d). The electroluminescence region is where the electric field intensity $E$ is high enough to excite Er ions, i.e. $qE\ell \geq \Delta$ with $q$ the unit charge, $\ell$ the mean free path of electrons and $\Delta$ the Er excitation energy from the ground to first excited state. Therefore, the width of the electroluminescence region is highly dependent on the reverse bias. According to the device principle of PN junction diode, the electric field intensity in the depletion region is spatially linear in both n and p region, reaching a maximum $E_{max}$ at the interface as shown in Figure 2(e). The electroluminescence width $L_{ex}$, represented dashed line in Figure 2(e) from $-L_{exp}$ to $L_{exn}$, can be expressed as eq.(1). It should be noted that here we used a relatively large voltage range to provide a large depletion region width, so as to obtain a sufficient number of $L_{ex}$ for fitting. However, such a large voltage range is not necessary during the measuring of output power and internal quantum efficiency in the following text.

$$L_{ex} = L_{exn} + L_{exp} = W_{dep}\frac{E_{max}-E_{min}}{E_{max}} \quad (1)$$

in which $E_{max} = \gamma\sqrt{(V_{bi} + V_R)}$ and $E_{min} = \gamma\sqrt{(V_{bi} + V_{min})}$ with $V_{bi}$ is the built-in potential, $V_R$ is the reverse bias voltage of the diode, $E_{min}$ is the minimum electric field required to excite Er ions by impact excitation and $V_{min}$ is the corresponding reverse bias voltage. The ratio $\gamma$ is written as $\gamma = \sqrt{2qN_aN_d/(\varepsilon_s(N_a + N_d))}$ with $q$ being the unit charge, $\varepsilon_s$ the dielectric constant of Si, $N_a$ and $N_d$ the net acceptor and donor concentration, respectively. From Figure 2(d), we find the full width at half maximum (FWHM) of the electroluminescence spatial profile at different reverse bias. Indeed, as shown in Figure 2(f), eq.(1) can well fit the electroluminescence FWHM that is dependent on reverse bias, from which we extract the net acceptor concentration $N_a$ as $2.0 \pm 0.15 \times 10^{15}$ cm$^{-3}$ and the minimum reverse bias $V_{min}$ as $1.87 \pm 0.46$ V. The p-doping concentration $N_a$ is consistent with what we found from resistivity measurements (see SI section I).

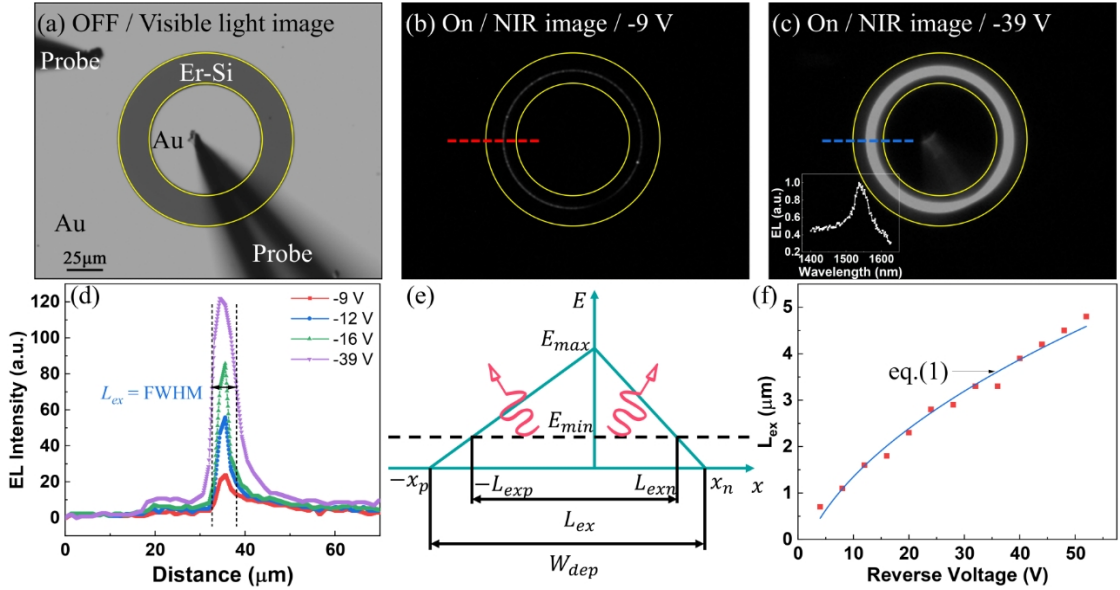

Figure 2. (a) Visible light image of LED. (b) and (c) NIR image of LED in (a) under reverse bias of -9 V and -39 V, respectively. Inset in (c) is the electroluminescence spectrum. (d) Spatial distribution of electroluminescence intensity along the red and blue dashes in (b) and (c), additional voltages are also provided. (e) Schematic of the relationship between emission region and electric field. (f)The dependence of emission region width on reverse voltage. The dots and solid line are experimental data and fitted data by eq.(1), respectively.

We place the LED in a Dewar vessel with infrared transparent window, cooled by liquid Nitrogen. Figure 3(a) shows the IV characteristics of the LED at different temperatures. The reverse current can be fitted with an empirical equation $I_R = I_0(e^{aV_R} - b)$. The reverse current decreases as the temperature lowers from room temperature, indicating that the reverse current is closely related to the thermal generation via SRH generation-recombination process[21]. To accurately measure the absolute power of LEDs, we use a commercial LED (L12509-0155G, Hamamatsu) with an emission angle similar to our LED to calibrate the light collection efficiency (0.2%) of the measurement system (see SI section II). This collection efficiency is also confirmed by the measurements in an integrating sphere (50 mm diameter, Thorlabs). Figure 3(b) shows the dependence of output photon flux on the electron flux under reverse voltage at difference temperature. The photon flux is converted from the power density, which is defined as the absolute power intensity divided by the electroluminescence area (the circular strip in Figure 2(c)). The internal quantum efficiency (IQE) calculated from Figure 3(b) is shown in Figure 3(c). It is worth noting that only about 2.3% of the photons can emit from bulk silicon into free space due to the high refractive index of Si[15, 22]. This efficiency is used to convert external quantum efficiency to internal quantum efficiency. At low reverse bias, most Er ions are in ground state and hot electrons can excite multiple Er ions by impact excitation. Therefore, the photon flux grows super linearly with the electron flux. As the reverse voltage increases to a certain point, most Er ions have been excited in the excitation state, as a result of which the emission flux starts to saturate. During this process, IQE first increases to the peak and then declines rapidly, as shown in Figure 3(c). At lower temperature, the thermal generation of electrons via Er related defects in Si bandgap is suppressed and IQE therefore increases. At 78 K, IQE reaches a record of 1.84%.

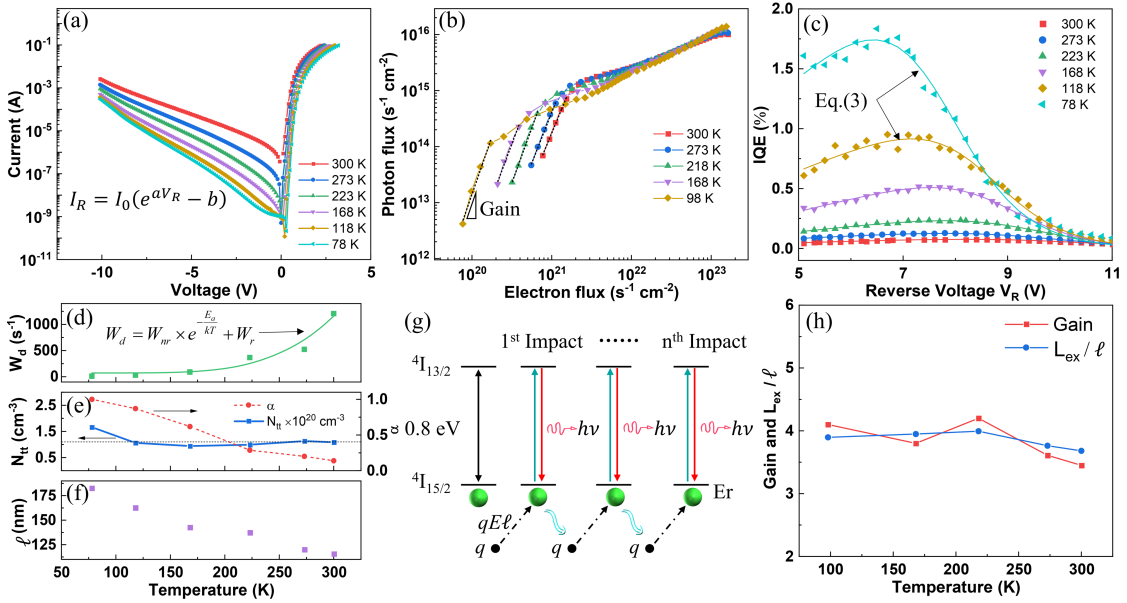

Figure 3. (a) IV curves at different temperatures. (b) Output photon flux vs. pump electron flux at different temperatures. (c) IQE vs. reverse voltage at different temperatures. All the solid lines are fitted using eq.(3). (d-f) Extracted relaxation rate $W_d$, optically active Er concentration $N_{tt}$, factor $\alpha$ and mean free path $\ell$ dependent on temperature. (g) Schematic of multiple impact excitation by one electron. (h) Gain and the ratio of $L_{ex}/\ell$ at different temperature. Gain is extracted from the slope indicated in the black dashes in (b).

The aforementioned Er concentration limiting process can be described by eq.(2) below in which $W_i \cong W_i' n\sqrt{qE\ell - \Delta}$ is the impact excitation probability for each Er ions for a given concentration $n$ of electrons with $W_i'$ being the coefficient determined by the factors such as the band structure of silicon and the Er ion size[15]. The kinetic energy is denoted as $qE\ell$, where $q$ is the unit charge, $E$ is the electric field intensity, $\ell$ is the mean free path of electrons and $\Delta$ is the Er excitation energy (0.8 eV) from ground to first excitation sate. $N_{tt}$ represents the concentration of all optically active Er ions, $N_{Er}$ is the concentration of Er ions in excited state and $W_d$ is the relaxation rate of Er ions in excitation state including radiative and nonradiative relaxation.

$$W_i(N_{tt} - N_{Er}) = W_d N_{Er} \qquad (2)$$

Suppose a fraction $\alpha$ of the relaxation rate $W_d$ is radiation emission. The emission rate $r_e = \alpha W_d N_{tt} L_{ex} A_c W_i/(W_d + W_i)$ with $L_{ex}$ being the width of electroluminescence region and $A_c$ the cross-section area perpendicular to radial direction at the n⁻ and p⁻ interface. The internal quantum efficiency IQE is given by the emission rate divided by the number of injection electron per second ($I_R/q$) as shown in eq. (3).

$$IQE = \frac{r_e}{I_R/q} = \frac{\alpha q W_d N_{tt} L_{ex} A_c W_i}{I_R(W_d + W_i)} \qquad (3)$$

The analytical IQE expression given in eq.(3) can well fit the experimental IQE at different temperature in Figure 3(c), after plugging into eq.(3) with $L_{ex}$ given in eq.(1), $W_i \cong W_i' n\sqrt{qE\ell - \Delta}$ taken from our previous publication [15] and the empirical reverse current $I_R$ found from Figure 3(a). The full IQE expression is given in SI section III. From the fittings, we extracted the relaxation rate $W_d$, the term $\alpha N_{tt}$, the mean free path $\ell$ and the minimum voltage for impact excitation $V_{min}$. All these parameters are dependent on temperature. According to our previous work and other literature, Er ions in excited state can be relaxed via temperature dependent nonradiative relaxation and temperature independent radiative emission $W_r$, i.e. $W_d = W_{nr} \exp\left(-\frac{E_a}{kT}\right) + W_r$ with $W_{nr}$ being the coefficient of nonradiative relaxation and $E_a$ the activation energy for back transfer and $kT$ the thermal energy. This analytical equation can nicely fit the experimental relaxation rate $W_d$ as shown in Figure 3(d) from which we extract $W_{nr}$ as $6.5 \times 10^5$ s$^{-1}$, $E_a$ as 170 meV and $W_r$ as 73 s$^{-1}$. These extracted parameters are comparable to what we found optically and other reports in literature[8, 23-25]. The fraction of radiative emission $\alpha$ is plotted in Figure 3(e) (right y axis). The concentration of optically active Er ions can be found by dividing the extracted term $\alpha N_{tt}$ with the $\alpha$ data in Figure 3(e). This calculation indicates that the optically active concentration of Er ions is ~$1.14 \times 10^{20}$ cm$^{-3}$, independent of temperature. The mean free path $\ell$ also increases from 115 nm at 300 K to 185 nm at 78 K as shown in Figure 3(f), which is not surprising, because a lower temperature will reduce the phonon scatterings.

Finally, let us revisit Figure 3(b) in which the emission flux is a function of injection electron flux. Interestingly, the derivative of the photon emission flux with respect to the electron flux at low end is significantly larger than 1. It implies that every additional electron injected into the diode excites multiple photons. This emission gain phenomenon likely occurs when the width of the electroluminescence region is longer than the electron mean-free path, i.e. $L_{ex} > \ell$. In this case, the electrons will be re-accelerated to collide with the Er ions after losing their kinetic energy in their previous colliding, as shown in Figure 3(g). Logically, the emission gain should be approximately equal to $L_{ex}/\ell$, i.e. G = $L_{ex}/\ell$. To validate this emission gain mechanism, we plot the average gain and the ratio of $L_{ex}/\ell$ as a function of temperature in Figure 3(h). The gain and the ratio are indeed comparable over the entire temperature range. (Note that although the mean-free path $\ell$ is larger at lower temperature in Figure 3f, the gain G does not increase at lower temperature because $L_{ex}$ will increase as $E_{min}$ decreases to meet $qE_{min}\ell = \Delta$, see Figure 3e.) We also repeat our measurements on similar LEDs with different radius. The correlation of G = $L_{ex}/\ell$ is persistently observed on these LEDs (see SI section IV). More interestingly, in our previous publication[15], the Er doped Si LED has a much higher carrier concentration in the p and n region. The electroluminescence region is as narrow as 10 nm and the mean-free path extracted from the theoretical fitting is ~ 4 nm. Consistently, we observed an experimental photon emission gain ~ 2.8 by impact excitation, very close to the ratio of the electroluminescence region width to the mean-free path (10 / 4 = 2.5). These observations further validate the fact that the photon emission gain is induced by multiple impact excitations of each electron as the electron transports through the high electric field depletion region.

## Conclusion

By leveraging reverse-biased Er-doped Si LEDs, we successfully achieved photon emission gain through impact excitation. Combined theoretical and experimental analysis demonstrate that widening the depletion region and reducing carrier concentrations significantly enhance Er excitation efficiency, while cryogenic operation suppresses non-radiative recombination, boosting IQE to 1.84%. Parameter fitting revealed the optically active Er concentration (~$10^{20}$ cm$^{-3}$) and temperature-dependent electron mean free path (115 nm to 185 nm). The designed wide depletion region enables single electrons to excite multiple Er ions, directly validating the gain mechanism G = $L_{ex}/\ell$. These findings not only deepen the understanding of Er-related luminescence mechanisms in silicon but also establish critical technological foundations for developing high-efficiency light sources in integrated silicon photonic quantum systems. Future studies may focus on optimizing doping processes and device architectures to advance applications in quantum communication and optical interconnects.

## Acknowledgement

This work was financially supported by the National Science Foundation of China (NSFC) (No. 62304131, W2412118, 92065103, 62305354), the special-key project of Innovation Program of Shanghai Municipal Education Commission (No. 2019-07-00-02-E00075) and the Oceanic Interdisciplinary Program of Shanghai Jiao Tong University (No. SL2022ZD107). The fabrication and measurement of LED was supported by Advanced Electronic Materials and Devices (AEMD) and Instrumental Analytical Center (IAC) of Shanghai Jiao Tong University. The Er and O implantation was done by the Ion Beam Center (IBC) at Helmholtz-Zentrum Dresden-Rossendorf (HZDR) and the B implantation was performed at Shanghai Institute of Technical Physics, Chinese Academy of Science.

semiconductors, Semiconductor Science and Technology, 8, 718 (1993).